\documentclass[onecolumn,11pt]{article}
\usepackage{amsmath}
\usepackage{amsfonts}
\usepackage{amssymb}
\usepackage{graphicx}

\newcommand{\ket}[1]{|#1\rangle}
\newcommand{\be}{\begin{equation}}
\newcommand{\ee}{\end{equation}}
\newcommand{\bq}{\begin{quote}}
\newcommand{\eq}{\end{quote}}
\newcommand{\bit}{\begin{itemize}}
\newcommand{\eit}{\end{itemize}}

\title{Why the wave function, of all things?}
\author{Ulrich Mohrhoff\\
\textit{\small Sri Aurobindo International Centre of Education}\\ 
\textit{\small Pondicherry 605002 India}\\
\ttfamily{\small ujm@auromail.net}}
\date{}
\normalsize
\begin{document}
\maketitle
\begin{abstract}
There are reasons to doubt that making sense of the wave function (other than as a probability algorithm) will help with the project of making sense of quantum mechanics. The consistency of the quantum-mechanical correlation laws with the existence of their correlata is demonstrated. The demonstration makes use of the fact (which is implied by the indeterminacy principle) that physical space is not partitioned ``all the way down,'' and it requires that the eigenvalue--eigenstate link be replaced by a different interpretive principle, whose implications are explored.
\end{abstract}

\vskip0.25in\noindent The question I am asking here is this:  why focus on the wave function?%
\footnote{An earlier version of this paper was contributed to an online workshop on the meaning of the wave function (http://www.ijqf.org/groups-2/meaning-of-the-wave-function/). A synopsis in the form of three posts introducing the paper is included in the Appendix.}
After all, there are a considerable number of formulations of quantum theory---Heisenberg's matrix formulation, Schr\"odinger's wave-function formulation, Feynman's path-integral formulation, the density-matrix formulation,\break Wigner's phase-space formulation, etc.---and only one of them uses wave functions. Clearly, making sense of the wave function and beating sense into the theory are distinct objectives, and there are strong reasons to doubt that pursuing the former will help with the latter.

While Bohr had insisted that the preparation of a system and the subsequent measurement of a system observable constitute a holistic phenomenon, which cannot be dissected into the unitary evolution of a wave function and a subsequent collapse of the same, today the task of beating sense into the theory is widely understood to be that of accounting for the collapse of the wave function associated with a measurement. That this line of attack is a red herring follows from two insolubility proofs of the so-called objectification problem \cite{Mittelstaedt,BLM}. These proofs assume that there is such a thing as a measurement process, and that this takes place in three steps: the state preparation, a continuous dynamical process called ``premeasurement''  ($p$), and the appearance of an outcome called ``objectification'' ($o$):
\begin{displaymath}
\sum_{k}c_k\ket{A_0}\ket{q_k}\stackrel{(p)}{\longrightarrow}
\sum_{k}c_k\ket{A_k}\ket{q_k}\stackrel{(o)}{\longrightarrow}
\ket{A(q)}\ket{q}.
\label{trip}
\end{displaymath}
The initial state assigns probability 1 to the outcome of a measurement which indicates that the apparatus is ready, and the final state assigns probability 1 to the outcome of a measurement which indicates that the apparatus indicates the outcome~$q$. To make the initial state represent the \emph{fact} that the apparatus is in the neutral state, and to make the final state represent the \emph{fact} that the apparatus indicates the outcome~$q$, one has to adopt the so-called eigenvalue--eigenstate link%
\footnote{Formulated thusly by Dirac \cite{Dirac}: ``The expression that an observable `has a particular value' for a particular state is permissible \dots\ in the special case when a measurement of the observable is certain to lead to the particular value, so that the state is an eigenstate of the observable.''}
(EEL) as an interpretive principle. This means that even if we had an explanation for the collapse transition  from $\sum_{k}c_k\ket{A_k}\ket{q_k}$ to $\ket{A(q)}\ket{q}$, it would not explain the objectification transition from probability~1 to factuality (``is'' or ``has''). All we can do is \emph{postulate}, via the EEL, that measurements have outcomes. 

But then we have a consistency problem. To be able to resolve it, we need to dispense with the EEL. Bub~\cite{Bub} has claimed that the unitary quantum dynamics can be made consistent with the existence of measurement outcomes by stipulating instead that ``the decoherence `pointer' selected by environmental decoherence'' is always definite. Decoherence then ``guarantees the continued definiteness or persistent objectivity of the macroworld.'' Decoherence, however, merely displaces the coherence of the system composed of apparatus and object system into the degrees of freedom of the environment, causing the objectification problem to reappear as a statement about the system composed of environment, apparatus, and object system. Since the mixture obtained by tracing out the environment does not admit an ignorance interpretation, it can resolve the problem only FAPP.

To demonstrate the consistency of the quantum-mechanical correlation laws with the existence of their correlata, which ought to be self-evident, we must desist from conceiving of the physical world as if it were directly accessible to our senses on all scales. In other words, we must not conceive of it as being spatially differentiated ``all the way down,'' which is what the aforementioned insolubility proofs implicitly assume. The demonstration has two parts: showing that quantum mechanics itself implies the incompleteness of the world's spatiotemporal differentiation, and deducing from this the existence of a non-empty class of objects whose positions are ``smeared out'' only relative to an imaginary spatiotemporal background that is more differentiated than the actual world \cite{Mohrhoff-QMexplained, ujm-opqf, Mohrhoff_manifesting, Mohrhoff-QS, Mohrhoff-book}.

\emph{Part 1.} Without the EEL, quantum mechanics can tell us that the probability of finding a particle in a \emph{given} region of space is~1, but it is incapable of  \emph{giving} us a region of space. For this a detector is needed. A detector is needed not only to indicate the presence of a particle in a region but also---and in the first place---to realize a region, so as to make it possible to attribute to the particle the property of being inside. Speaking more generally, a measurement apparatus is needed not only to indicate the possession of a property by a quantum system but also---and in the first place---to realize a set of properties so as to make them available for attribution to the system. This is vintage Bohr, and it couldn't be more true. But if detectors are needed to realize regions of space, space cannot be intrinsically partitioned.  If at all we conceive of it as partitioned, we can do so only as far as regions of space can be realized---i.e., to the extent that the requisite detectors are physically realizable. Because this extent is limited by the indeterminacy principle, the spatial differentiation of the objective world is incomplete.

\emph{Part 2.} In an incompletely differentiated world, there will be objects whose position distributions are and remain so narrow that there are no detectors with narrower position distributions. If anything truly deserves the label ``macroscopic,'' it is these objects. While decoherence arguments can solve the objectification problem only FAPP, they quantitatively support the existence of macroscopic positions---positions whose indefiniteness is never revealed in the only way it could be revealed, i.e., through a departure from what the classical  laws predict. The testable correlations between the outcomes of measurements of macroscopic positions are therefore consistent with \emph{both} the classical \emph{and} the quantum laws. This makes it possible to attribute to macroscopic positions a measurement-independent reality, and it enables them to define the obtainable values of observables and to indicate the outcomes of measurements.$\,\Box$

But is it enough to establish the consistency of the quantum-mechanical correlation laws with the reality of their correlata? Is there no ontological message? There is, but to uncover it we must get rid of the notion that the wave function's dependence on time is the continuous time-dependence of an evolving physical state. As (macroscopic) detectors are needed to realize attributable positions, so (macroscopic) clocks are needed to realize attributable times. The $t$ in $\psi(t)$ refers to the macroscopically realized and indicated time of the measurement to the possible outcomes of which $\psi(t)$ serves to assign probabilities.

Although a distinction has to be made between formulations and interpretations of the theory, the choice of a formulation cannot but bias the range of available interpretations. Let us now compare the wave-function formulation with Feynman's~\cite{FHS}. In both formulations we have two dominant dynamical principles. In the former they are unitary evolution and collapse. In the latter they are summation over amplitudes (followed by taking the absolute square of the sum) and summation over probabilities (preceded by taking the absolute square of each amplitude). From the former's point of view, unitary evolution is ``normal'' and hence not in need of explanation. What calls for explanation is collapse. From the latter's point of view, adding probabilities is ``normal'' inasmuch as it is what classical probability leads us to expect. What calls for explanation is why we have to add amplitudes whenever we are required to do so. 

The reason we postulate collapse in the context of the wave-function formulation is the indisputable factuality of measurement outcomes \emph{together} with the EEL, inasmuch as this provides the link between factuality and the immediate consequence of a collapse (probability~1 for a particular outcome). What, then, could be the reason we have to add amplitudes? In answer to this question I have proposed the following interpretive principle as a replacement for the untenable EEL \cite{Mohrhoff-QMexplained, ujm-opqf, Mohrhoff_manifesting, Mohrhoff-QS, Mohrhoff-book}:
\begin{itemize}
\item[(I)]Whenever quantum mechanics requires us to add amplitudes, the distinctions we make between the alternatives correspond to nothing in the physical world.
\end{itemize}
This is a statement about the structure of the physical world, not merely a statement of our practical or conceptual limitations. In one case we are thus stumped by the dual problems of collapse and objectification, while in the other case we know why ``[a]ny determination of the alternative taken by a process capable of following more than one alternative destroys the interference between alternatives''~\cite{FHS}. (By stating the indeterminacy principle in this way, Feynman \emph{et al.} do not mean to imply that the ``destruction'' is brought about by a physical process.)

Applied to a two-way interferometer experiment, the replacement (I) for the EEL tells us that the distinction we make between ``the particle went through the left arm'' and ``the particle went through the right arm'' corresponds to nothing in the physical world. Since this distinction rests on a distinction between regions of space, it follows that space cannot be an intrinsically differentiated expanse. Its so-called parts need to be physically realized, and we have seen that the indeterminacy principle prevents them from being realized ``all the way down.''

Applied to an elastic scattering event involving two particles of the same type (two incoming particles $N$ and $S$, two outgoing particles $E$ and $W$), the interpretive principle (I) tells us that the distinction we make between the alternatives

\medskip\centerline{$N\rightarrow E,S\rightarrow W$\quad and\quad $N\rightarrow W,S\rightarrow E$}

\medskip\noindent corresponds to nothing in the physical world. There is no answer to the question ``Which outgoing particle is identical with which incoming one?'' And why is that so? Because the incoming particles (and therefore the outgoing ones as well) are one and the same entity. What's more, there is no compelling reason to believe that this identity ceases when it ceases to have observable consequences owing to the presence of individuating properties.  We are free to take the view that \emph{intrinsically} each particle is numerically identical with every other particle; what presents itself here and now with these properties and what presents itself there and then with those properties is one and the same entity.%
\footnote{According to French \cite{SEP-French}, quantum mechanics is ``compatible with two distinct metaphysical `packages,' one in which the particles are regarded as individuals and one in which they are not.'' Esfeld~\cite{Esfeld2013} disagrees: it is not ``a serious option to regard quantum objects as possessing a primitive thisness (haecceity) so that permuting these objects amounts to a real difference.''}
In what follows I shall call it ``Being.''

Perhaps the main reason it is so hard to make sense of the quantum theory is that it answers a question we are not in the habit of asking. Instead of asking what the ultimate constituents of matter are and how they interact and combine, we should  ask: \emph{how are forms manifested?} To this question there is a straightforward answer \cite{Mohrhoff_manifesting,Mohrhoff-QS}: \emph{The shapes of things are manifested with the help of reflexive spatial relations.} By entering into reflexive spatial relations, Being gives rise to (i)~what \emph{looks like} a multiplicity of relata if the reflexive quality of the relations is ignored and (ii)~what \emph{looks like} a substantial expanse if the spatial quality of the relations is reified. As Leibniz said, \emph{omnibus ex nihilo ducendis sufficit unum}---one is enough to create everything from nothing. A single self-existent entity is enough to create both the relata we call particles and the expanse we call space.

The following brief reflection leads to the same conclusion. While the non-relativistic theory allows us to conceive of a physical system as being composed of a definite number of parts, and to conceive of its form as being composed of a definite number of spatial relations (to which values can be attributed only if and when they are measured), the relativistic theory requires us to treat the number of a system's parts as just another quantum observable, which has a definite value only if and when it is measured. There is therefore a clear sense in which a quantum system is always one, the number of its parts having a definite value only if and when it is measured.

To my mind, the most fruitful way to understand the necessary distinction between the classical or macroscopic domain (which contains measure\-ment-independent properties) and the non-classical or quantum domain (whose properties exist only if, when, and to the extent that they are measured) is that it is essentially a distinction between the \emph{manifested world} and its \emph{manifestation}. The curious mutual dependence of the two domains, pointed out by Landau and Lifshitz%
\footnote{``[Q]uantum mechanics \dots\ contains classical mechanics as a limiting case, yet at the same time it requires this limiting case for its own formulation''~\cite{LL77}.}
and by Redhead,%
\footnote{``In a sense the reduction instead of descending linearly towards the elementary particles, moves in a circle, linking the reductive basis back to the higher levels''~\cite{Redhead1990}.}
is then readily understood. The manifestation of the world consists in a transition from a condition of complete indefiniteness and indistinguishability to a condition of complete or maximal definiteness and distinguishability, and what occurs in the course of this transition---what is not completely definite or distinguishable---can only be described in terms of probability distributions over what is completely definite and distinguishable. What is instrumental in the manifestation of the world can only be described in terms of its result, the manifested world.%
\footnote{It is worth noting here that the indeterminism of quantum mechanics is rooted in an underlying indeterminacy. Instead of consisting fundamentally in the existence of unpredictable changes disrupting a predictable evolution, it is a consequence of  indeterminacies that evince themselves through unpredictable transitions in the values of outcome-indicating positions (Bub's ``decoherence pointers'').}

Quantum mechanics thus presents us with a so far unrecognized kind of causality---unrecognized, I believe, within the scientific literature albeit well-known to metaphysics, for the general philosophical pattern of a single world-essence manifesting itself as a multiplicity of physical individuals is found throughout the world.%
\footnote{Some of its representatives in the Western hemisphere are the Neoplatonists, John Scottus Eriugena, and the German idealists. The quintessential Eastern example is the original (pre-illusionist) Vedanta of the Upanishads \cite{Phillips, SA_Isha, SA_Kena}.}
This causality is associated with the atemporal process of manifestation, which effects the transition from complete indefiniteness and indistinguishability to complete or maximal definiteness and distinguishability. It must be distinguished from its familiar temporal cousin, which links states or events across time or spacetime. The latter plays no role in the manifestation. Being part of the world drama, it does not take part in setting the stage for it.%
\footnote{Ladyman and Ross \cite{LadyRoss} concur: ``the idea of causation has similar status to those of cohesion, forces, and [individual] things. It is a concept that structures the notional worlds of observers\dots.  There is no justification for the neo-scholastic projection of causation all the way down to fundamental physics and metaphysics.''}

The atemporal causality%
\footnote{A participant of the workshop asked: ``Would you say that this new type of causality is effectively happening in some new dimension of time?'' I would not. Yet while the transition from a condition of complete indefiniteness and indistinguishability to a condition of complete or maximal definiteness and distinguishability (via emergent stages populated by numerically identical particles, non-visualizable atoms, and partly visualizable molecules) is neither temporal nor spatial, we cannot help conceive it in temporal terms, just as we cannot help conceive temporal relations in spatial terms, as aspects of a 4-dimensional continuum. Although this spatialization of time fails to do justice to the qualitative aspects of our  experience of time (change and succession), we would be hard pressed to deal with the relativistic interdependence of distances and durations without conceiving of time as if it were another spatial dimension. Quantum mechanics presents us with a similar Catch-22: although by temporalizing that transition we fail to do justice to its nature, we would be hard pressed to envision it without temporalizing it.}
associated with the process of manifestation casts new light on quantum theory's mysterious violation of remote outcome-independence---the fact that the probability of a measurement outcome in one laboratory can depend on the measurement outcome obtained in another laboratory, even if the spatiotemporal relation between the two measurements is spacelike.%
\footnote{The diachronic correlations between events in timelike relation actually are as mysterious as the synchronic correlations between events in spacelike relation. While we know how to calculate either kind of correlation, and therefore know how to calculate the probabilities of possible events on the basis of actual events, we know as little of a physical process by which an event here and now contributes to determine the probability of a \emph{later} event \emph{here} as we know of a physical process by which an event here and now contributes to determine the probability of a \emph{distant} event \emph{now}. All we have is correlations.}
The reason why local explanations do not work may be the same as the reason why the manifestation of the spatiotemporal world cannot be explained by processes that connect events \emph{within} the spacetime arena. The manifestation of the world is the nonlocal event \emph{par excellence}. Instead of being an event \emph{in} spacetime, it is, depending on one's point of view, either ``outside'' of spacetime (i.e., not localized at all) or coextensive with spacetime (i.e., completely delocalized). It is the process by which Being enters into reflexive relations and matter and space come into being as a result. It is the transition by which Being acquires both the aspect of a multiplicity of relata (if the reflexive quality of the relations is ignored) and the aspect of a substantial expanse (if the spatial quality of the relations is reified). The atemporal causality of this transition supports the folk causality that connects objects across space and events across spacetime, which helps us make sense of the manifested world as well as of the cognate world of classical physics, but which throws no light on the process of manifestation nor on the quantum correlations that are instrumental in the process.

\bigskip\noindent\textbf{\Large Appendix}

\medskip\noindent Prior to the workshop a synopsis of the paper was provided in three parts/ posts, which are reproduced here.

\medskip\noindent\textbf{Part 1.} The question I am asking here is this:  why focus on the wave function? After all, there are (by a recent count) at least nine different formulations of quantum theory. The wave-function formulation presents us with two mysteries: why is the unitary evolution disrupted by the occasional collapse, which results in the assignment of probability 1 to a particular outcome, and why is probability 1 sufficient for the factuality of that outcome? The only way the latter mystery can be solved is by adopting the so-called eigenvalue-eigenstate link, which postulates that probability 1 is sufficient for the factuality of that outcome. But doing so makes the quantum-mechanical correlation laws, which presuppose the existence of correlata (measurement outcomes), inconsistent with the existence of the correlata, which is absurd.

The consistency of the quantum-mechanical correlations with the existence of their correlata can be demonstrated if one gives up the eigenvalue--eigenstate link. The demonstration takes place in two steps. First I show (here in outline, in greater detail in some of the referenced papers) that the world is not spatially differentiated (or partitioned) ``all the way down'': its spatial (and hence spatiotemporal) differentiation is incomplete. From this I deduce the existence of a non-empty class of objects whose positions are ``smeared out'' only relative to an imaginary spatiotemporal background that is more differentiated spacewise than the actual world. If anything truly deserves the label ``macroscopic,'' it is these objects. The testable correlations between the outcomes of measurements of their positions are consistent with both the classical and the quantum laws. This makes it possible to attribute to these positions the measurement-independent reality that is lost by giving up the eigenvalue--eigenstate link, and it enables them to define the obtainable values of observables and to indicate the outcomes of measurements.

Trigger phrases like ``measurement'' and ``macroscopic object'' are likely to elicit accusations of instrumentalism, Copenhagenism, or some such. Common or garden instrumentalism, however, leaves the meaning of ``macroscopic'' up for grabs. What is accomplished here is a consistent definition of ``macroscopic'' in the theory's own terms. And that's only the beginning.

\medskip\noindent\textbf{Part 2.} The eigenvalue--eigenstate link is an interpretive principle that saves the appearances in the context of the wave-function formulation of quantum mechanics. To go beyond a metaphysically sterile instrumentalism, a different interpretive principle needs to be used, as well as as a different formulation of quantum mechanics: Feynman's. Both the wave-function formulation and Feynman's feature a pair of dynamical principles; in the former they are unitary evolution and collapse, in the latter they are summation over amplitudes and summation over probabilities. In the context of the wave-function formulation, unitary evolution seems ``normal''; what calls for explanation is collapse. In the context of Feynman's formulation, adding probabilities seems ``normal''; what calls for explanation is why we have to add amplitudes. What is at issue, then, is not what causes the wave function to collapse but why we have to add amplitudes whenever quantum mechanics requires us to do so. To answer this question I have proposed the following interpretive principle:

(I) \emph{Whenever quantum mechanics requires us to add amplitudes, the distinctions we make between the alternatives correspond to nothing in the physical world. They cannot be objectified (represented as real).}

Next, I apply this interpretive principle to two paradigmatic setups, one concerning distinctions between regions (of space or spacetime), the other concerning distinctions between things. Applied to a two-way interferometer experiment, (I)  tells us that the distinction we make between ``the particle went through the left arm'' and ``the particle went through the right arm'' corresponds to nothing in the physical world, whence it follows that physical space cannot be an intrinsically differentiated expanse. Its so-called parts need to be physically realized by the sensitive regions of detectors (defined in terms of macroscopic positions), and the indeterminacy principle prevents them from being realized ``all the way down.''

Applied to an elastic scattering event involving two particles of the same type (two incoming particles N and S, two outgoing particles E and W), (I) tells us that the distinction we make between the alternative identifications (N=E, S=W) and (N=W, S=E) corresponds to nothing in the physical world. There is no answer to the question ``Which outgoing particle is identical with which incoming one?''

\medskip\noindent\textbf{Part 3.} So why is there no answer to the question ``Which outgoing particle is identical with which incoming one?''? Because the incoming particles (and therefore the outgoing ones as well) are one and the same entity. What's more, there is no compelling reason to believe that this identity ceases when it ceases to have observable consequences owing to the presence of individuating properties.  We are free to take the view that intrinsically each particle is numerically identical with every other particle; what presents itself here and now with these properties and what presents itself there and then with those properties is one and the same entity. For want of a better word I call it ``Being'' with a capital~B.

As I see it, the main reason it is so hard to make sense of the quantum theory is that it answers a question we are not in the habit of asking. Instead of asking what the ultimate constituents of matter are and how they interact and combine, we should  ask: how are forms manifested? This question, too, has a straightforward answer: The shapes of things are manifested with the help of reflexive spatial relations. By entering into reflexive spatial relations, Being gives rise to (i) what looks like a multiplicity of relata if the reflexive quality of the relations is ignored and (ii) what looks like a substantial expanse if the spatial quality of the relations is reified.

To my mind, the most fruitful way to understand the necessary distinction between the classical or macroscopic domain (which contains measure\-ment-independent properties) and the non-classical or quantum domain (whose properties exist only if, when, and to the extent that they are measured) is that it is essentially a distinction between the manifested world and its manifestation. 

Quantum mechanics thus presents us with a so far unrecognized kind of causality---unrecognized within the scientific literature albeit well-known to metaphysics. This causality is associated with the atemporal process of manifestation, which effects the transition from a condition of complete  indefiniteness and indistinguishability to a condition of maximal definiteness and distinguishability. It must be distinguished from its familiar spatiotemporal cousin, which links states or events across time or spacetime. The latter causality plays no role in the manifestation, which is why it is inapplicable to the subject-matter of quantum mechanics---the correlation laws that are instrumental in the process of manifestation. The atemporal causality associated with the process of manifestation thus casts new light on quantum theory's mysterious violation of outcome-independence. The reason why local explanations do not work is the same as the reason why the manifestation of the spatiotemporal world cannot be explained by processes that connect events within the spacetime arena.

\end{document}